\documentstyle[twoside,fleqn,espcrc2,epsf]{article}

\newcommand{\AmS}{{\protect\the\textfont2
  A\kern-.1667em\lower.5ex\hbox{M}\kern-.125emS}}

\def\cpt{$\chi$PT }
\def\qcpt{q$\chi$PT }

\catcode`@=11
\def\@sim#1#2{\setbox0=\hbox{$\sim$}\lower.9\ht0\vbox{\baselineskip0pt
              \lineskip0.1ex\ialign{$\m@th#1\hfill##\hfill$\crcr#2\crcr
              \sim\crcr}}}
\def\lsim{\mathrel{\mathpalette\@sim<}}
\def\gsim{\mathrel{\mathpalette\@sim>}}
\catcode`@=12

\title{
\vspace*{-35pt}
{\normalsize \hfill {\sf UTCCP-P-52}} \\
\vspace*{-6pt}
{\normalsize \hfill {\sf Oct.\ 1998}} \\
Recent Results from the CP-PACS Collaboration}

\author{
R.~Burkhalter\address{Center for Computational Physics,
University of Tsukuba, Tsukuba, Ibaraki 305, Japan}
for the CP-PACS Collaboration\thanks{CP-PACS Collaboration: 
S.~Aoki, G.~Boyd, R.~B., S.~Eji\-ri, M.~Fukugita, S.~Hashimoto,
Y.~Iwasaki, K.~Kanaya, T.~Kaneko, Y.~Kuramashi, K.~Nagai, M.~Okawa,
H.~P.~Sha\-nahan, A.~Ukawa, T.~Yoshi\'e}
}

\begin{document}

\begin{abstract}
We present an overview of recent results from the CP-PACS computer on the
quenched light hadron spectrum and an on-going two-flavour full QCD study.
We find that our quenched hadron mass results are compatible with the 
mass formulae predicted by quenched chiral perturbation theory, 
which we adopt in our final analysis. 
Quenched hadron masses in the continuum limit show unambiguous and
systematic deviations from experiment.  
For our two-flavour full QCD simulation we present preliminary results on
the light hadron spectrum, quark masses and the static potential.
The question of dynamical sea quark effects in these 
quantities is discussed.
\end{abstract}

\maketitle

\section{Introduction}

Calculating the light hadron spectrum from first principles has been one of
the main objectives of lattice QCD since its
formulation\cite{ref:spectrum-review}.  
Pioneering attempts towards this goal in the quenched
approximation\cite{ref:quench-pioneer} were followed by an increase of
understanding and control of various systematic uncertainties due to finite
lattice spacing, finite size 
and the problem of chiral extrapolations.
An outstanding effort carried out by the GF11 collaboration\cite{ref:GF11}
led to a quenched spectrum agreeing within 5--10\% with experiment.
With its history quenched QCD can therefore be considered as a
matured field where an additional increase of calculational accuracy will
finally tell us how the quenched spectrum deviates from experiment. 

Last year the CP-PACS collaboration presented first results of
an effort in this direction\cite{ref:CPPACS_quench}. The simulation has
been completed in the meantime and the final results have been presented at
this conference\cite{ref:yoshie}.

Progress in the full QCD spectrum study has been comparatively slower. Even
though full QCD simulations have a history almost as long as
quenched ones\cite{ref:full-history}, 
the field is still at an exploring stage. 
In particular how in detail the spectrum
is influenced by the presence of light sea quarks has not yet been fully 
answered\cite{ref:guesken}.  Recently the CP-PACS collaboration undertook a
further step in the direction of a realistic full QCD simulation. At this
conference first preliminary results have been 
presented\cite{ref:kanaya,ref:kaneko,ref:hugh,ref:taniguchi}.

In this article we present an overview of the two calculations of the CP-PACS 
collaboration described above. 
In the first part we discuss in detail our final results for the quenched 
hadron spectrum, devoting some length to the issue of chiral extrapolations. 
In the second part we switch to full QCD
where we present our main results, focussing on whether
effects from sea quarks can be seen.

\section{Simulation of Quenched QCD with the Standard Wilson Action}

We list the parameters of our quenched simulation in 
Table~\ref{tab:quench-params}, where the partition size of the CP-PACS 
computer employed and the execution time for one cycle of configuration
generation and hadron propagator calculations are also given for each
lattice size.  

The runs are performed on four lattices covering the lattice spacing in the
range $a^{-1} \!=\! 2$--4~GeV for extrapolation to the
continuum limit. The physical lattice size in the spatial direction is
chosen to be around $La \!\approx\! 3$~fm where finite size effects are
expected to be negligible. Gauge configurations are generated with
over-relaxation and heat bath algorithms mixed in a 4:1 ratio.  

For each value of $\beta$ quark propagators are calculated for five quark 
masses corresponding to $m_{PS}/m_V$ of about 0.75, 0.7, 0.6, 0.5 and
0.4. The first two values are taken to be around the strange quark for
interpolation, and the last value is our attempt to control chiral 
extrapolations better than in previous spectrum studies with the Wilson quark
action.  We calculate hadron masses for equal as well as unequal quark 
masses so that we can determine the complete light hadron spectrum with
degenerate $u$ and $d$ quarks and a heavier strange quark.  

Since Lattice 97 we have increased the number of configurations from 91 to 150
at $\beta \!=\! 6.47$.  For technical details of the simulation we refer to 
\cite{ref:CPPACS_quench}. 

\begin{table}[tb]
\setlength{\tabcolsep}{0.2pc}
\newlength{\digitwidth} \settowidth{\digitwidth}{\rm 0}
\catcode`?=\active \def?{\kern\digitwidth}
\caption{Parameters of quenched simulation.
Values of $a^{-1}$ are determined from $m_\rho$ obtained by 
a quenched chiral perturbation theory fit to vector meson masses.}
\label{tab:quench-params} 
\begin{tabular}{llllllr}
\hline\hline
$\beta$ & size  &  $a^{-1}$  & $La$ & \#   & \#      & hrs/  \\
        &       &  [GeV]     & [fm] & conf & PU      & conf  \\
\hline
5.90 & $32^3\!\!\times\!\!56$  & 1.934(16) & 3.26(3) & 800 & \,\,\,256 &  3.0 \\
6.10 & $40^3\!\!\times\!\!70$  & 2.540(22) & 3.10(3) & 600 & \,\,\,512 &  4.8 \\ 
6.25 & $48^3\!\!\times\!\!84$  & 3.071(34) & 3.08(3) & 420 & 1024 &  6.8 \\ 
6.47 & $64^3\!\!\times\!\!112$ & 3.961(79) & 3.18(6) & 150 & 2048 & 15.5 \\ 
\hline\hline
\end{tabular}
\vspace{-15pt}
\end{table}

\section{Chiral Extrapolation in Quenched QCD}

Extrapolation to the chiral limit is a major nontrivial component of 
light hadron spectrum calculations\cite{ref:Gottlieb96}. 
In addition to the difficulty
that simulations are done far from the chiral limit there is the issue of the
choice of the functional form of hadron masses as a function of the quark
masses.  
In considering this issue it should be asked whether the
actual data supports the validity of quenched chiral perturbation theory
(q$\chi$PT)~\cite{ref:Sharpe_qChPT,ref:BG_qChPT,ref:BG_DecayRatio,ref:Booth,ref:LabrenzSharpe},
which predicts characteristic singularities in hadron masses in the chiral
limit. 
One also has to examine how results differ if different possible fitting 
functions are employed. 
In the following we examine these questions in detail.

\subsection{pseudoscalar meson sector}

The dependence of the pseudoscalar (PS) meson mass $m_{PS}$ on its two
valence quark masses $m$ and $m_s$ is predicted by \qcpt to have the
form~\cite{ref:BG_qChPT}
\begin{eqnarray}
m_{PS}^2 &=&  A(m_s+m)\left\{ 1 - \delta \left [\ln(2mA/\Lambda_\chi^2) 
              \right . \right . \nonumber \\
  &+& \left . \left . m_s \ln(m_s/m)/(m_s-m) \right ] \right\} \nonumber \\
  &+& B(m_s+m)^2  + C(m_s-m)^2 + \cdots \label{eq:mps-quench}
\end{eqnarray}

\begin{figure}[tb]
\vspace{-7mm}
\begin{center}
\leavevmode
\hspace*{-4mm}
\epsfxsize=8.2cm
\epsfbox{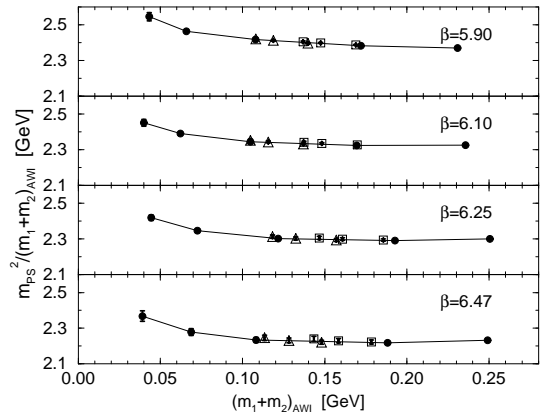}
\end{center}
\vspace{-66mm}
\caption{Ratio of $m_{PS}^2$ to quark mass from axial Ward
identity. Filled symbols are for degenerate quark masses, open symbols for
non-degenerate cases.}
\label{fig:PS-quark-quench}
\vspace{-15pt}
\end{figure}

To test whether our data is consistent with the presence of a logarithmic
term we plot in Fig.~\ref{fig:PS-quark-quench} the ratio of the PS meson 
mass squared to the quark mass. 
We employ quark masses determined from an
extended axial current Ward identity (AWI), 
$\nabla_\mu\!A^{\rm ext}_\mu\!=\!2m_q^{\rm AWI}a P$,
avoiding ambiguities due to
the determination of the critical hopping parameter.  
We see a clear increase of the ratio for small
quark masses in agreement with (\ref{eq:mps-quench}).

\begin{figure}[tb]
\vspace{-13mm}
\begin{center}
\leavevmode
\hspace*{-3mm}
\epsfxsize=8.2cm
\epsfbox{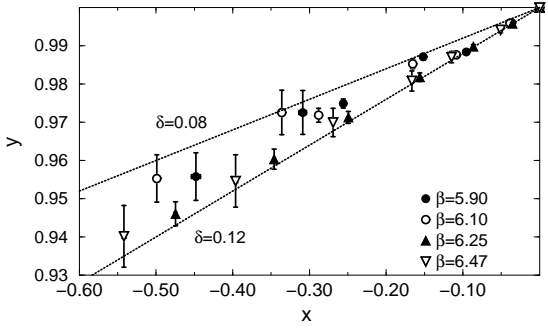}
\end{center}
\vspace{-73mm}
\caption{Pseudoscalar meson test for quenched chiral logarithms 
(see Eqs.~(\ref{eq:BGratio-x}) and (\ref{eq:BGratio-y})).}
\label{fig:BGtest-quench}
\vspace{-15pt}
\end{figure}

A more quantitative test can be made by forming the two variables
\begin{eqnarray}
x &=& 2 - \frac{m_s+m}{m_s-m}\log\left (\frac{m_s}{m}\right ),
      \label{eq:BGratio-x} \\
y &=& {\frac{2m}{m_s+m} \frac{m_K^2}{m_\pi^2}} \times 
   {\frac{2m_s} {m_s+m} \frac{m_K^2}{m_\eta^2}}, 
    \label{eq:BGratio-y} 
\end{eqnarray}
where $\pi$ and $\eta$ are degenerate pseudoscalar mesons with an equal
quark mass of  
$m$ or $m_s$, and $K$ is the non-degenerate meson. 
The two quantities are related by
\begin{equation}
y = 1 + \delta \cdot x,
\end{equation}
if the $O((m_s-m)^2)$ term in (\ref{eq:mps-quench}) is ignored.
In Fig.~\ref{fig:BGtest-quench} we plot our mass data in terms of $x$ 
and $y$ for two choices of $m_s$ ($m_{PS}/m_V=$ 0.75 or 0.7) 
and three for $m$ ($m_{PS}/m_V=$ 0.6, 0.5 or 0.4).  
All our data lie within a narrow wedge bounded by two lines corresponding
to $\delta \!=\! 0.08$--0.12.

The mass ratio $m_K^2/m_\pi^2$ originally proposed\cite{ref:BG_qChPT}  
has corrections due to both quadratic terms in (\ref{eq:mps-quench}). 
For this ratio we find that our data does not lie on a common line, 
indicating that the
inclusion of the term $O((m+m_s)^2)$ is important for the description of
our data.

\begin{figure}[tb]
\vspace{-13mm}
\begin{center}
\leavevmode
\hspace*{-2.5mm}
\epsfxsize=8.2cm
\epsfbox{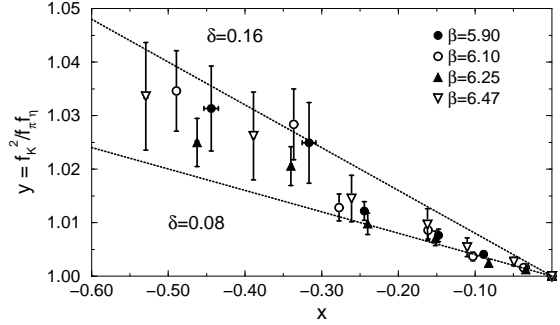}
\end{center}
\vspace{-73mm}
\caption{Decay constant ratio test.}
\label{fig:decaytest-quench}
\vspace{-15pt}
\end{figure}

A different test uses a ratio of decay
constants~\cite{ref:BG_DecayRatio}; 
the quantity $x$ from (\ref{eq:BGratio-x}) and 
$y\!=\! f_K^2/(f_\pi f_\eta)$  are related by $y\!=\! 1 \!-\! \delta \cdot
x/2$.  
As shown in Fig.~\ref{fig:decaytest-quench} the data lies again in a wedge,
bounded by two lines corresponding to $\delta \!=\! 0.08$--0.16.

Having seen evidence for quenched chiral logarithms in the above tests, we
fit the pseudoscalar meson mass with (\ref{eq:mps-quench}), taking into
account the 
correlation between meson masses at different quark masses.  
We drop the quadratic term proportional to $C$ since the mass test 
(\ref{eq:BGratio-x}) and (\ref{eq:BGratio-y})
suggests its effect to be small.  
The fits are done in terms of $1/K$ since it has no statistical error,
which means that we use the quark mass derived from the vector Ward 
identity (VWI), $m_q^{\rm VWI}a \!=\! (1/K-1/K_c)/2$. 

\begin{figure}[tb]
\vspace{-13mm}
\begin{center}
\leavevmode
\hspace*{-4mm}
\epsfxsize=8.2cm
\epsfbox{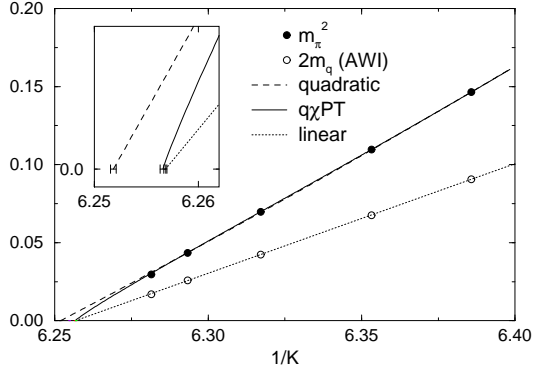}
\end{center}
\vspace{-67mm}
\caption{Chiral extrapolation of the PS meson and the AWI quark mass 
at $\beta\!=\!5.9$.}
\label{fig:chiral-ps-quench}
\vspace{-16pt}
\end{figure}

In Fig.~\ref{fig:chiral-ps-quench} we show the degenerate pseudoscalar
meson mass at 
$\beta\!=\!5.9$ as a function of $1/K$ together with fit curves 
from (\ref{eq:mps-quench}).  Results for a quadratic fit in $m_q^{\rm VWI}a$ 
(dashed lines),  and the AWI quark mass data with a linear fit (dotted line) 
are also plotted.  We observe that the AWI quark mass is well described by 
a linear function in $1/K$.  This means that a proportionality 
$m_q^{\rm VWI}\!\propto\!m_q^{\rm AWI}$ holds between the two definitions of 
quark mass, which in turn justifies our use of different quark
mass definitions for mass tests and fitting. 
Another important feature is a good agreement of the critical hopping
parameter $K_c$ from the q$\chi$PT fit of pseudoscalar meson masses and
from a linear  
fit of AWI quark masses, whereas they differ by as much as $17\sigma$ if a 
quadratic fit is employed  for the pseudoscalar meson masses.

\begin{figure}[tb]
\vspace{-15mm}
\begin{center}
\leavevmode
\hspace*{-4mm}
\epsfxsize=8.2cm
\epsfbox{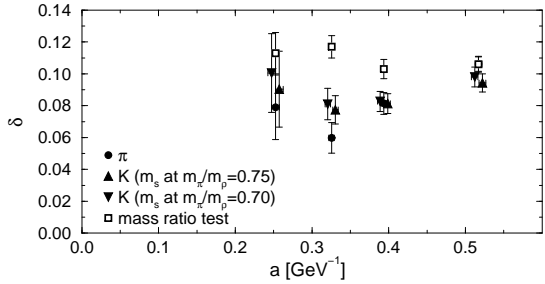}
\end{center}
\vspace{-77mm}
\caption{Parameter $\delta$ from fit to pseudoscalar meson masses (filled
symbols) 
and mass test (open symbols).
For clarity the data is slightly shifted in the $x$-direction.}
\label{fig:delta-quench}
\vspace{-10pt}
\end{figure}

Results for the parameter $\delta$ obtained from the fit 
are shown in Fig.~\ref{fig:delta-quench} as function of the
lattice spacing for degenerate and non-degenerate data separately.  
Values obtained with a linear fit of the mass
test in Fig.~\ref{fig:BGtest-quench} are also shown (open squares). 
At each value of $a$, the results of independent fits to
degenerate and non-degenerate masses agree with each other within error
bars.  The values from the mass test are somewhat higher in comparison. 
Overall however, all the results for $\delta$ are mutually in reasonable 
agreement. 

To summarize our examinations in the pseudoscalar sector, we find the two
different 
direct tests to provide a positive evidence for the presence of \qcpt
logarithms.   The value of $\delta$ is in the range 
$\delta \approx 0.10(2)$.

\subsection{vector meson sector}

For vector meson masses q$\chi$PT gives\cite{ref:Booth} 
\begin{eqnarray}
m_{\rho} & = & m_V^0 + C_{1/2}  m_{\pi} + 2C_1 m_{\pi}^{2},
\label{eq:Rho-quench} \\
m_{K^*}  & = & m_V^0 \nonumber \\
         & + & \frac{C_{1/2}}{6} \left\{ \frac{3}{2} 
               (m_{\pi} + m_{\eta}) + 
               2 \frac{m_{\eta}^3-m_{\pi}^3}{m_{\eta}^2-m_{\pi}^2} \right\} 
               \nonumber \\
         & + & C_1 (m_{\pi}^2 + m_{\eta}^2) + O(m_{PS}^3).  
\label{eq:Kst-quench}
\end{eqnarray}
where $C_{1/2}\!=\!- 4\pi g_2^2 \delta$ with $g_2$ a phenomenological 
coupling constant of the vector meson quenched chiral Lagrangian.

An examination of mass ratios such as 
$(m_\rho-m_\phi)/(m_\pi-m_\eta)=C_{1/2}+O(m_{PS})$ calculated for our data 
indicates a small negative value in the chiral limit.  
The evidence, however, is less compelling than for pseudoscalar mesons
because a 
chiral extrapolation is needed in order to extract $C_{1/2}$.

As an alternative analysis we attempt to fit data with the predicted forms 
(\ref{eq:Rho-quench},\ref{eq:Kst-quench}). 
To examine whether our data exhibits the full pattern of dependence on
pseudoscalar 
masses we employ a simultaneous fit to degenerate and non-degenerate
masses.
Such a combined fit involves a large number of data points, for which we find 
only uncorrelated fits to be practically feasible. 

\begin{figure}[tb]
\vspace{-13mm}
\begin{center}
\leavevmode
\hspace*{-5mm}
\epsfxsize=8.5cm
\epsfbox{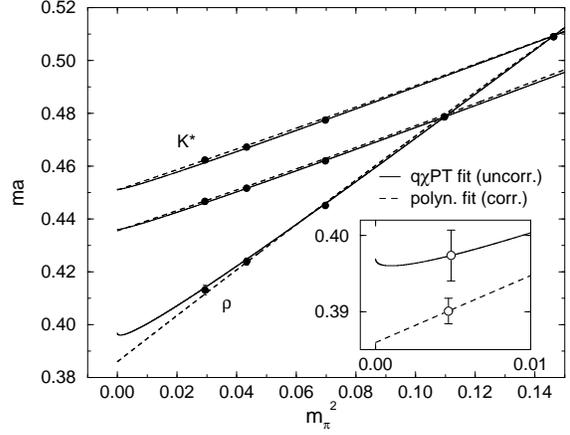}
\end{center}
\vspace{-60mm}
\caption{Chiral extrapolation of vector meson masses at
$\beta\!=\!5.9$. Open symbols in the inset are extrapolated values 
at physical quark mass.}
\label{fig:chiral-vec-quench}
\vspace{-10pt}
\end{figure}

This fit reproduces our vector meson mass data well as illustrated in 
Fig.~\ref{fig:chiral-vec-quench}.  Clearly visible in the fit curve for 
$m_\rho$ (solid line) is the effect of the $O(m_{PS})$ term, 
giving rise to an increase close to the chiral limit.   
The magnitude of the effect, however, is small, and this holds for all
$\beta$.   
Correspondingly, small negative values 
for $C_{1/2}$ are found from the fits (see Table~\ref{tab:quench-c12}), 
consistent with indications from mass ratio tests.
The average of $C_{1/2}$ over $\beta$ gives $C_{1/2}\!=\! -0.071(8)$, 
which is ten times smaller compared to the
phenomenological estimate $C_{1/2}\!=\!- 4\pi g_2^2 \delta \!\approx\! -
0.71$ (for $g_2 \!=\! 0.75$ and $\delta \!=\! 0.1$).

\subsection{baryon sector}

For the baryons we find a situation for mass ratio tests less 
conclusive than for vector mesons. 
Therefore we follow the same strategy and fit masses of
different degeneracies simultaneously. We assume \qcpt mass
formulae~\cite{ref:LabrenzSharpe}, i.e. for octet baryons
\begin{eqnarray}
   m_{\Sigma} &=& m_O^0  \nonumber \\
              &+& \left (4F^2w_{uu} - 4(D-F)Fw_{us} \right . \nonumber \\
              & & + \left . (D-F)^2w_{ss} \right )/2 \nonumber \\
              &-& 4 b_F m_{uu}^2 + 2(b_D-b_F) m_{ss}^2 \nonumber \\
              &+& (2D^2/3 - 2F^2) v_{uu} \nonumber \\ 
              &+& (2D^2/3 - 4DF + 2F^2) v_{us}, \\ 
   m_{\Lambda}&=& m_O^0 \nonumber \\
              &+& \left ( (4D/3-2F)^2w_{uu} + (D/3+F)^2w_{ss} \right . 
                  \nonumber \\
              & &  - \left . 2(4D/3-2F)(D/3+F)w_{us} \right )/2 \nonumber \\
              &+& 4 (2b_D/3-b_F)m_{uu}^2 - 2(b_D/3+b_F)m_{ss}^2 \nonumber \\
              &+& (2D^2/9 - 8DF/3 + 2F^2) v_{uu} \nonumber \\
              &+& (10D^2/9 - 4DF/3 - 2F^2) v_{us}, 
\end{eqnarray}
and for decuplet baryons
\begin{eqnarray}
   m_{D}      &=& m_D^0  \nonumber \\
              &+& 5H^2/162 \, (4w_{uu} + 4w_{us} + w_{ss}) \nonumber \\
              &+& C^2/18 \, (w_{uu} - 2w_{us} + w_{ss})   \nonumber \\
              &+& c \left (2m_{uu}^2 + m_{ss}^2 \right ) ,
\end{eqnarray}
where $m_{uu}$ and $m_{ss}$ are $\pi$ and $\eta$ meson masses and
$w_{us} = -2\pi\delta (m_{ss}^3-m_{uu}^3)/(m_{ss}^2-m_{uu}^2)$, 
$v_{us} = (m_{us}^3/16\pi f^2)$ and $f$ is the pion decay constant.

\begin{figure}[tb]
\vspace{-9mm}
\centerline{\epsfxsize=8.5cm \epsfbox{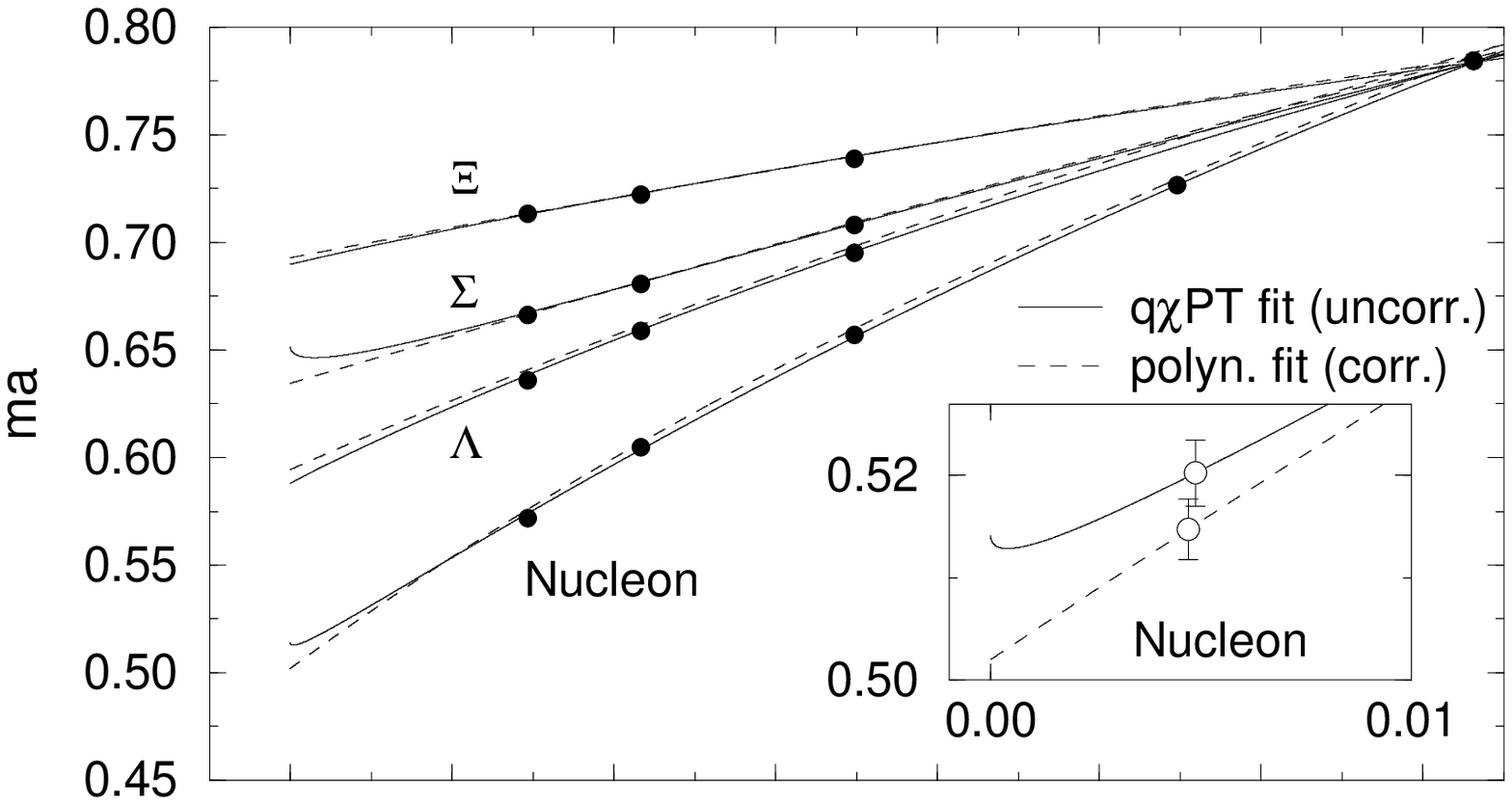}}
\vspace{-78mm}
\centerline{\epsfxsize=8.5cm \epsfbox{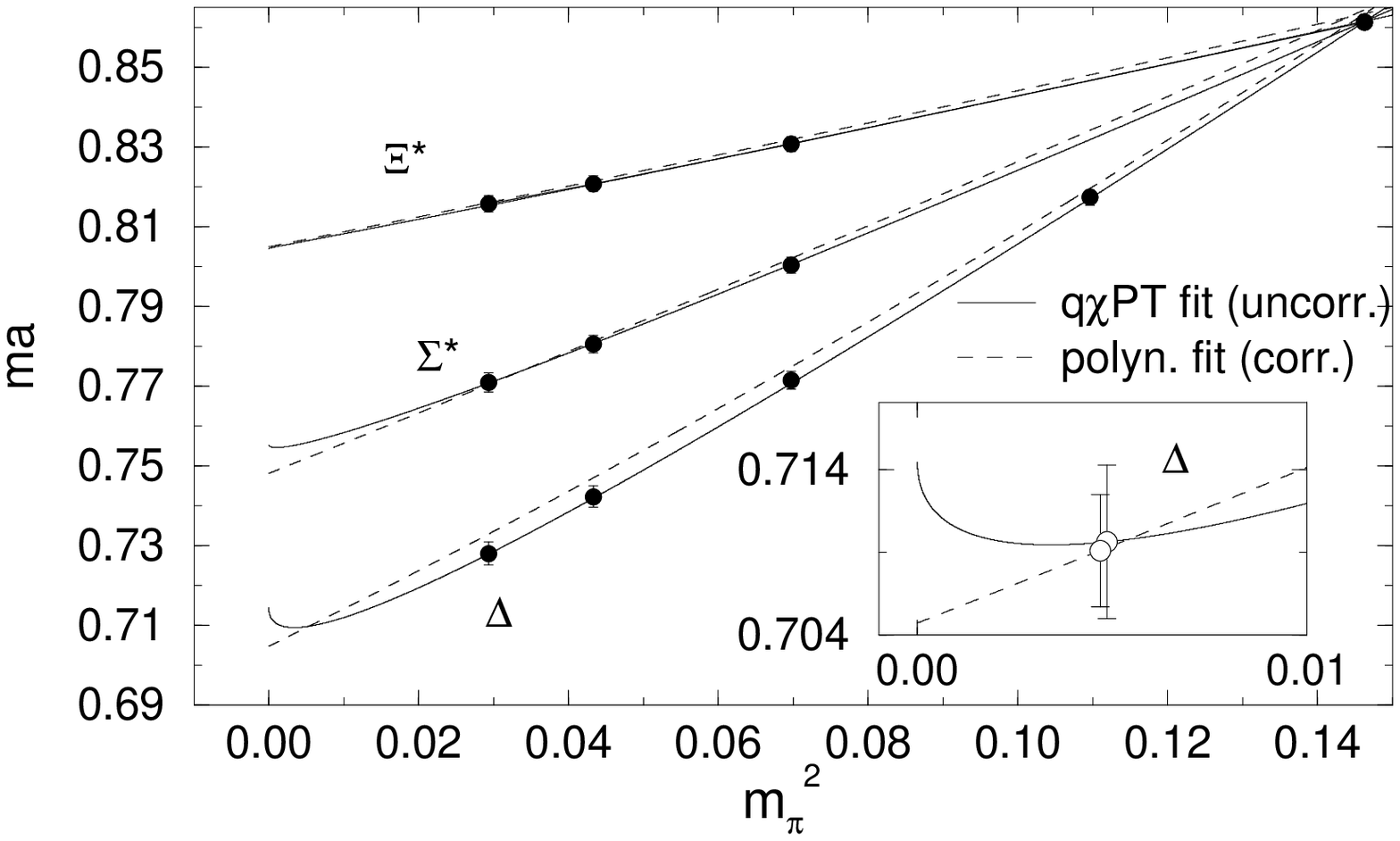}} 
\vspace{-70mm}
\caption{Chiral extrapolation of baryon masses at
$\beta\!=\!5.9$. Open symbols in insets are extrapolated values at the
physical quark mass.}  
\label{fig:chiral-bar-quench}
\vspace{-15pt}
\end{figure}

These forms are slightly simplified compared to the full 
formula in order to get stable fits. 
For decuplet baryons all leading linear terms in $m_{PS}$ 
and all quadratic terms are included, but not the cubic terms $O(m_{PS}^3)$.
For octet baryons, on the other hand, a bending in the nucleon mass data 
necessitates a retention of such terms; we keep those arising from the
octet-octet 
couplings $F$ and $D$, but not those due to the 
decuplet-octet coupling $C$. 
We also note that $f\!=\!93$~MeV and $\delta\!=\!0.1$ are fixed in the fits.

The result of the mass fits to octet and decuplet baryons is shown at
$\beta\!=\!5.9$ in Fig.~\ref{fig:chiral-bar-quench}. The increase of masses
close 
to the chiral limit as a quenched artifact is visible in the fit curves 
for the nucleon, $\Sigma$, $\Delta$ and $\Sigma^*$, but is small compared to 
the error after extrapolation.  
For the nucleon the coefficient of the $O(m_{PS})$ term 
$C_{1/2} \!=\! - 3\pi (D\!-\!3F)^2\delta/2$ obtained from the fits has an
average value $-0.118(4)$ (see also Table~\ref{tab:quench-c12}). This is
considerably smaller than a phenomenological estimate $-0.27$ for
which $\delta \!=\! 0.1$, and $F \!=\! 0.5$, $D \!=\! 0.75$ estimated from
experiment\cite{ref:VBernard} are used. The contribution of this term to
the nucleon mass is therefore about 2\% or 15~MeV. 
A similar value is found for $\Delta$.

\begin{table}[tb]
\setlength{\tabcolsep}{0.06pc}
\caption{
Coefficient $C_{1/2}$ of the $m_{PS}$ term for vector meson and
baryons.}
\label{tab:quench-c12} 
\begin{tabular}{l|llll}
\hline\hline
         & $\beta=5.90$  & $\beta=6.10$  & $\beta=6.25$ & $\beta=6.47$ \\
\hline
$\rho$   & $-0.058(27)$  & $-0.075(28)$ & $-0.065(35)$ & $-0.155(72)$  \\
$N$      & $-0.126(15)$  & $-0.120(15)$ & $-0.116(13)$ & $-0.087(30)$   \\ 
$\Delta$ & $-0.169(33)$  & $-0.145(39)$ & $-0.101(40)$ & $-0.184(104)$  \\ 
\hline\hline
\end{tabular}
\vspace{-17pt}
\end{table}

Summarizing, our data does not provide direct evidence for the validity of 
\qcpt in vector meson and baryon sectors. 
However, \qcpt formulae describe the overall pattern of the data well, 
with the fit parameters taking reasonably constant values at different 
values of $\beta$.  Our data is therefore compatible with the presence of 
$O(m_{PS})$ terms.  Their magnitudes are smaller than expected, amounting 
to about 2\% of vector and baryon masses.

\subsection{polynomial chiral fits}
\label{sec:polynomial}

\begin{figure}[tb]
\vspace{-4mm}
\centerline{\epsfxsize=8.5cm \epsfbox{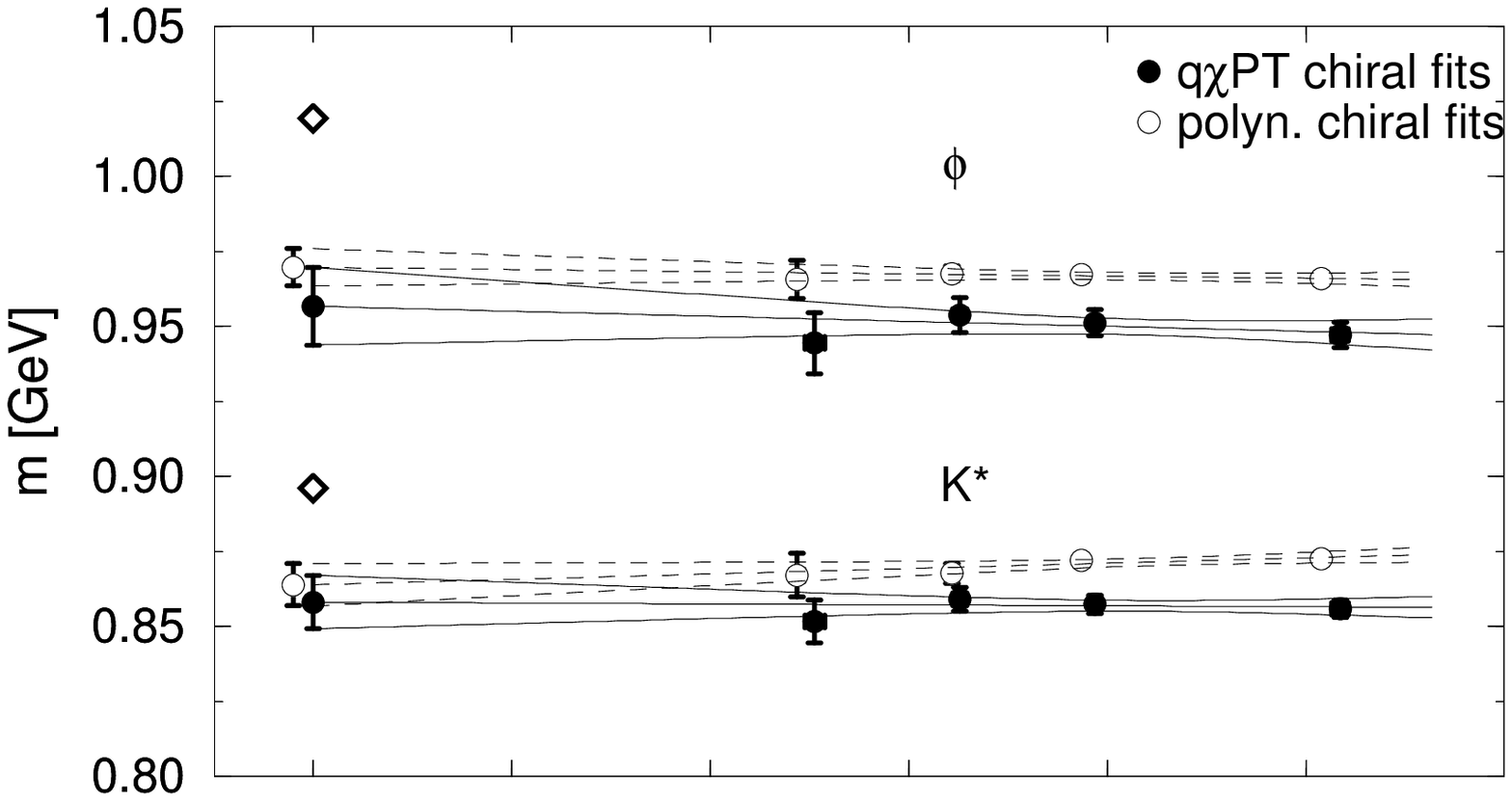}}
\vspace{-9.1cm}
\centerline{\epsfxsize=8.5cm \epsfbox{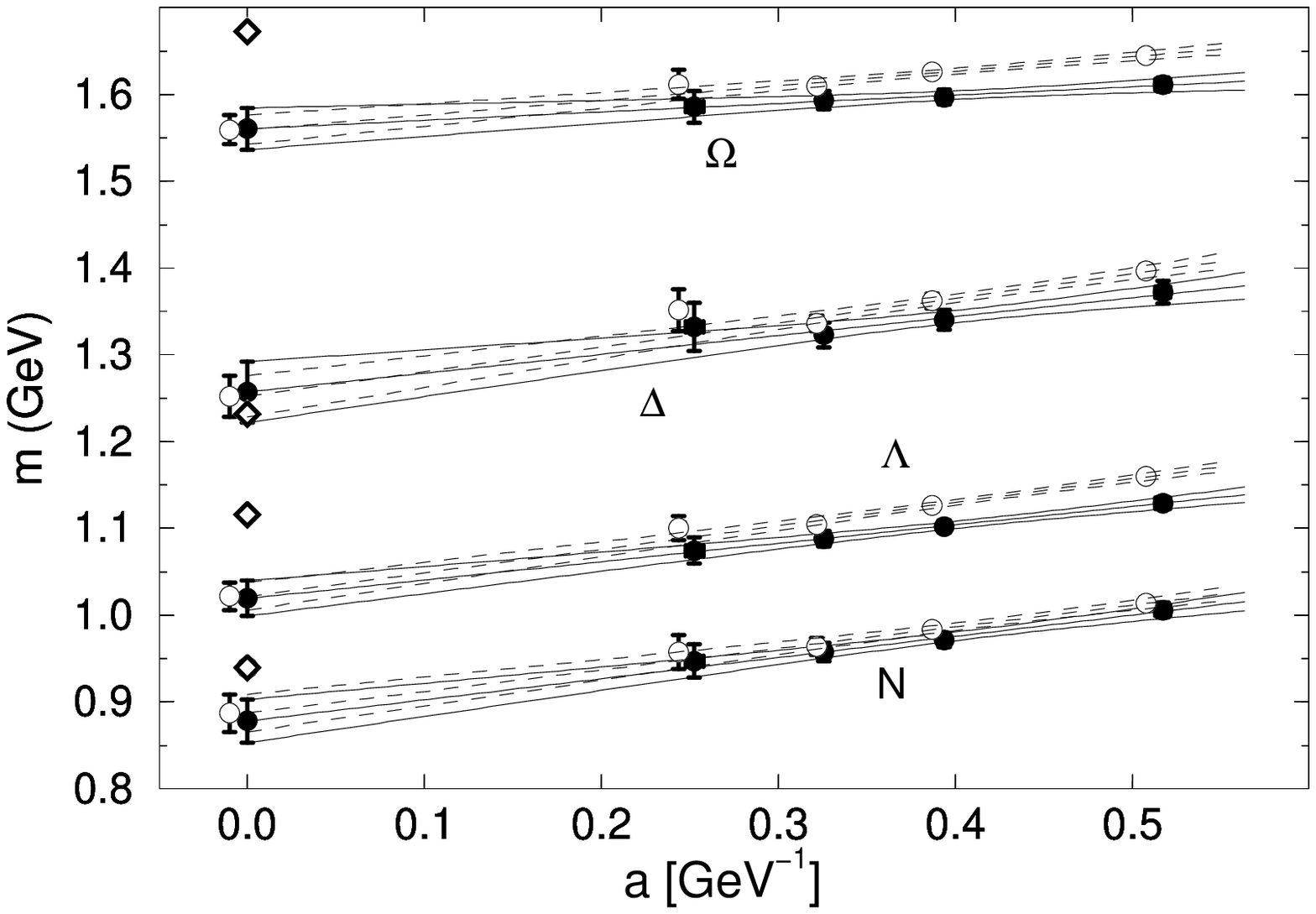}} 
\vspace{-5.8cm}
\caption{Continuum extrapolation of quenched hadron masses 
with $M_K$ as input for \qcpt and polynomial chiral fits.
}
\label{fig:Cont-quench}
\vspace{-15pt}
\end{figure}

In order to investigate the effect of choosing 
different fitting functions
we have also made chiral extrapolations employing polynomials in $1/K$
(cubic for $N$, quadratic otherwise). Hadrons of different degeneracies are
fitted independently and correlations between different quark masses are
taken into account. The results of these fits are already plotted for each 
channel in Figs.~\ref{fig:chiral-ps-quench}, \ref{fig:chiral-vec-quench} and
\ref{fig:chiral-bar-quench} by dashed lines.
The continuum extrapolations with \qcpt and polynomial fits are compared in 
Fig.~\ref{fig:Cont-quench} for meson masses and typical baryon masses.

The two types of fits do lead to different results after chiral extrapolation 
at each $\beta$, which can be as significant as $5\sigma$ in the largest case. 
In absolute magnitude, however, this difference corresponds to 2--3\% 
at most.  Furthermore, after continuum extrapolation, the results from
polynomial fits do not differ beyond 1.5\% from the values obtained with
\qcpt fits, which is a $1.5\sigma$ effect. 

We conclude that \qcpt and polynomial chiral extrapolations, while 
possibly leading to sizably different continuum values in principle, 
actually yield results which do not differ beyond a 2\% level.

\section{Quenched Light Hadron Spectrum}

Based on the results of analyses described in Sec.~3 we choose 
chiral extrapolations using \qcpt formulae for the calculation
of our final spectrum results.
The physical point for the degenerate $u$ and $d$ quarks is fixed by
$M_{\pi}$(135.0) and $M_{\rho}$(768.4), and we use $M_K$(497.7) or
$M_{\phi}$(1019.4) for the $s$ quark. The lattice scale is set with
$M_{\rho}$.  

\begin{figure}[tb]
\vspace{-14mm}
\begin{center}
\leavevmode
\hspace*{-5mm}
\epsfxsize=8.5cm
\epsfbox{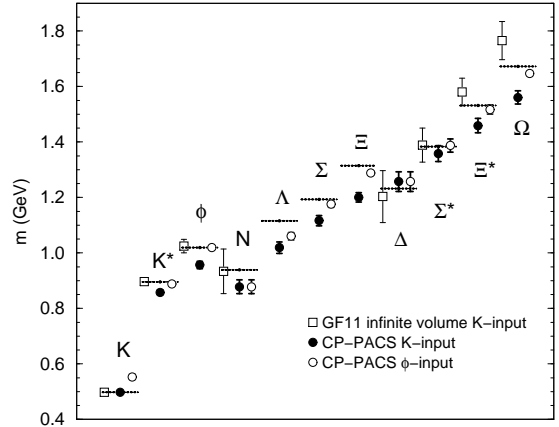}
\end{center}
\vspace{-60mm}
\caption{Final results of the quenched light hadron spectrum in the
continuum limit. Experimental values and results from
GF11~\cite{ref:GF11} are plotted for comparison.} 
\label{fig:Spectrum-quench}
\vspace{-15pt}
\end{figure}

We observe in Fig.~\ref{fig:Cont-quench} that the continuum extrapolation 
is well described by a linear function in the lattice spacing.  
Indeed, if we parameterize the $a$-dependence as
$m\!=\!m_0(1+\alpha a)$, we find the typical scale $\alpha$ to be
$\alpha\!\sim\!0.2$~GeV.  Hence higher order terms are expected to be
small, {\it e.g., } $(\alpha a)^2\!\sim\!0.01$ at 
$a\!=\!0.5\,{\rm GeV}^{-1}$.  We thus employ a linear continuum extrapolation. 
We present our final result for the quenched light hadron spectrum 
in Fig.~\ref{fig:Spectrum-quench}.

This figure shows that the quenched light hadron spectrum systematically 
deviates from experiment.  If one uses $M_K$ as input to fix the $s$ quark 
mass, the mass splitting between $K$ and $K^*$ ($\phi$) in the meson sector 
is underestimated by 9.5\% (12\%), which is a 
4.3$\sigma$ (4.9$\sigma)$ effect.  
In the baryon sector, the octet masses are smaller, by 7.6\% (4.5$\sigma$)
on the average, and the decuplet mass splittings are 30\% (1.1$\sigma$ for
individual level spacings, but 3.2$\sigma$ for the $\Delta-\Omega$
splitting) short of the experimental splitting.
With $M_\phi$ as input, the $K^*$ meson mass is in agreement with 
experiment, and the discrepancies in the baryon masses are reduced. 
However, the $K-K^*$ meson hyperfine splitting is still smaller by 
16\% which has a statistical significance of 6.1$\sigma$.

These conclusions remain unchanged compared to those presented at 
Lattice 97~\cite{ref:CPPACS_quench}, 
where we employed a linear chiral extrapolation in $1/K$ (cubic for $N$ and
quadratic for $\Lambda$), except in numerical details for individual channels. 
The most conspicuous change is the decrease of the nucleon mass away from 
the experimental value, and that of the $\Delta$ mass toward it.  Strange
baryon masses  
with $M_K$ as input have also decreased.  The changes, however, are at most
1.5$\sigma$ of the error of our final results, and do not affect the overall 
pattern of deviation between the quenched and experimental spectrum.

\section{Quark Masses in Quenched QCD}

\begin{figure}[tb]
\vspace{-18mm}
\begin{center}
\leavevmode
\hspace*{-5mm}
\epsfxsize=8.5cm
\epsfbox{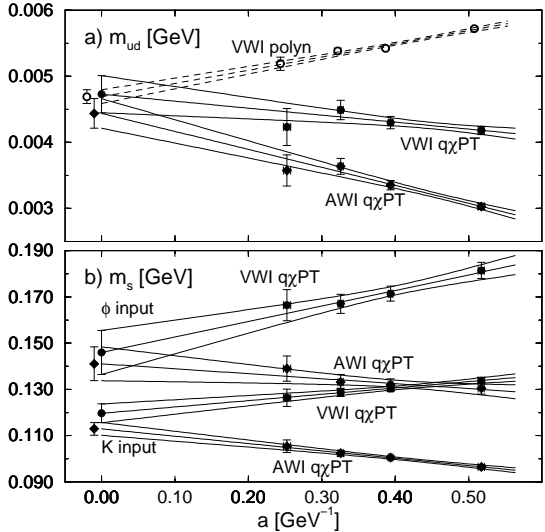}
\end{center}
\vspace{-40mm}
\caption{Continuum extrapolation of light quark masses in the
$\overline{\rm MS}$ scheme at $\mu=$ 2 GeV.} 
\label{fig:quark-quench}
\vspace{-15pt}
\end{figure}

The use of \qcpt fits to pseudoscalar meson masses has important
consequences for VWI quark masses. As already noted with 
Fig.~\ref{fig:chiral-ps-quench} the critical hopping parameter is different
for polynomial and \qcpt fits of pseudoscalar meson masses.  
This translates into a different estimate of
$m_q^{\rm VWI}$, influencing the average $u$ and $d$ quark mass most.  

In Fig.~\ref{fig:quark-quench} we show the quark mass results as a function 
of the lattice spacing. 
The conclusion, reported at Lattice 97, that the VWI and AWI quark masses
converge in the continuum limit remains unchanged also when \qcpt fits are
used. Another point to note is that the continuum result of the VWI quark 
masses obtained with quadratic polynomial chiral fits of 
pseudoscalar meson masses are consistent within error bars with those of
\qcpt fits,  
which is similar to the situation with hadron masses discussed in
Sec.~\ref{sec:polynomial}. 

For our final values of the quark masses we make a combined linear continuum
extrapolation of the VWI and AWI quark masses from \qcpt fits together, 
restricting them to a
common value in the continuum limit (see Fig.~3 in \cite{ref:yoshie}). 
We obtain $m_{ud}\!=\! 4.6(2)$~MeV and $m_s\!=\! 
115(2)$~MeV ($M_K$ input) or $m_s\!=\! 143(6)$~MeV ($M_\phi$ input) in the
$\overline{\rm MS}$ scheme at $\mu=$ 2 GeV.
The value of $m_{ud}$ is 12\% larger than the estimate reported at Lattice
97, where linear chiral extrapolations of pseudoscalar meson masses have
been used.

\section{Pseudoscalar Meson Decay Constants}

\begin{figure}[tb]
\vspace{-17mm}
\begin{center}
\leavevmode
\hspace*{-3mm}
\epsfxsize=8.2cm
\epsfbox{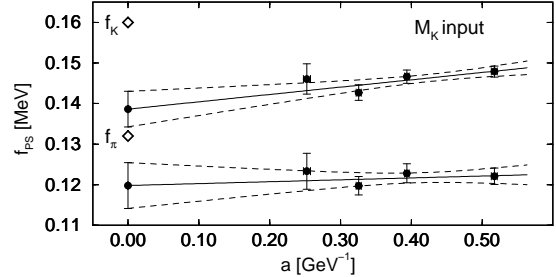}
\end{center}
\vspace{-73mm}
\caption{Pseudoscalar meson decay constants.}
\label{fig:ps-decay-quench}
\vspace{-15pt}
\end{figure}

PS meson decay constants are determined from the local axial
current. Because the quark mass dependence of degenerate decay constants
has not yet been determined within q$\chi$PT, we employ a quadratic
polynomial chiral extrapolation for $f_\pi$ and a linear extrapolation for
$f_K$. The results as a function of the lattice spacing are shown in
Fig.~\ref{fig:ps-decay-quench}.
We obtain $f_\pi=$ 120(6) MeV and $f_K=$ 139(4) MeV in the continuum limit,
which are 10 and 15\% smaller than experiment, respectively. The ratio
$f_K/f_\pi-1 =$ 0.156(29) is also smaller than the experimental value 0.212.

\section{Full QCD Simulation}

Having observed systematic deviations of the quenched hadron spectrum from
experiment, the natural question to ask is whether the deviation 
can be explained as a consequence of the effect of dynamical quarks.
Due to the large increase of computer time required for simulating full QCD
compared to the quenched approximation, we are forced to work at coarse
lattice spacings where $a^{-1} \!\lsim\! 2$~GeV, and are therefore led to
employ improved actions. 

In an initial study \cite{ref:CPPACS_full} we
compared the scaling behaviour of various combinations of improved gauge
and quark actions in full QCD at lattice spacings $a^{-1} \!\approx\!
1$~GeV.
Based on the outcome of this comparison we decided to employ for gauge
fields the RG improved action consisting of plaquette and 1$\times$2
rectangular Wilson loop proposed by Iwasaki \cite{ref:Iwasaki83}. For
quarks we use the SW clover action \cite{ref:clover} with mean-field
improved value of $c_{SW}$: $c_{SW}=P^{-3/4}$ with $P =
1-0.8412\cdot\beta^{-1}$ the one-loop value of the plaquette.

Simulations are carried out with two flavours of sea quarks, to be
identified with the degenerate $u$ and $d$ quarks, while the strange and
heavier quarks are treated in the quenched approximation. In
Table~\ref{tab:full-params} we give an overview of our run parameters.
Three lattice spacings in the range $a^{-1}\!\approx\!1$--2~GeV are used
for continuum extrapolation.
Lattice sizes are chosen so that the spatial size is kept around $La
\!\approx\!  2.4$~fm (except for $\beta\!=\!2.2$ where it turned out to be
smaller).
For chiral extrapolations, simulations are performed at four sea quark
masses corresponding to $m_{PS}/m_V \approx 0.8$, 0.75, 0.7 and
0.6.

\begin{table}[tb]
\setlength{\tabcolsep}{0.2pc}
\caption{Parameters of full QCD simulation.}
\label{tab:full-params} 
\begin{tabular}{ll|llll}
\hline\hline 
size          & $a^{-1}$ [GeV] & $K_{\rm sea}$ & $m_{PS}$
& \#traj  & time     \\
\#PU          & $La$ [fm]      &               & $\;/m_V$             
&         &  $\;$ /traj   \\
$\beta$       &                &               &              
&         &  $\;$ [hrs]  \\
$\rm c_{SW}$  &                &               &              
&         &          \\
\hline\hline 
$12^3\!\!\times\!\!24$ & 0.917(10) & .1409 & .8060(7) & 6250 & 0.11 \\
64                     & 2.58(3)   & .1430 & .753(1)  & 5000 & 0.16 \\
1.80                   &           & .1445 & .696(2)  & 7000 & 0.27 \\
1.60                   &           & .1464 & .548(4)  & 5250 & 0.93 \\
\hline 
$16^3\!\!\times\!\!32$ & 1.288(15) & .1375 & .8048(9) & 7000 & 0.11 \\
256                    & 2.45(3)   & .1390 & .751(1)  & 7000 & 0.16 \\
1.95                   &           & .1400 & .688(1)  & 7000 & 0.26 \\
1.53                   &           & .1410 & .586(3)  & 5000 & 0.85 \\
\hline 
$24^3\!\!\times\!\!48$ & 2.45(9)   & .1351 & .800(2)  & 1250 & 0.44 \\
512                    & 1.93(7)   & .1358 & .752(3)  & 1350 & 0.70 \\
2.20                   &           & .1363 & .702(3)  & 1610 & 0.99 \\
1.44                   &           & .1368 & .637(6)  & 1265 & 1.98 \\
\hline
\hline
\end{tabular}
\vspace{-15pt}
\end{table}

Configurations are generated with the HMC algorithm. To speed up our code
we have implemented several improvements. 
For the integration of molecular dynamics equations we employ the scheme
\begin{displaymath}
T_P(\frac{b}{2}d\tau) T_U(\frac{d\tau}{2})
T_P((1\!-\!b)d\tau )
T_U(\frac{d\tau}{2}) T_P(\frac{b}{2}d\tau), 
\end{displaymath}
where $T_P(d\tau)$ moves the gauge field and $T_U(d\tau)$ the conjugate
momenta by a step $d\tau$. This scheme has errors of the same order as
the standard leapfrog 
but the main contribution 
to the error 
is removed by the choice $b \!=\! (3-\sqrt{3})/3$~\cite{ref:SexWein}. For
our heaviest quarks $d\tau$ can be taken about a factor 3 larger than for
leapfrog, leading to a 30\% gain in computer time. The gain decreases for
decreasing quark masses and is lost at our smallest quark mass. For the
inversion of the quark matrix we use the even/odd preconditioned BiCGStab
algorithm~\cite{ref:bicgstab}. Its performance compared to MR increases
with decreasing quark mass. We observe a gain of about 40\% at
$m_{PS}/m_V \!\approx\! 0.7$. CPU times needed for the creation of one
unit of trajectory and numbers of processors are listed in
Table~\ref{tab:full-params}.

The light hadron spectrum is calculated at each sea quark mass with valence
quark masses corresponding to $m_{PS}/m_V \approx 0.8$, 0.75,
0.7, 0.6 and 0.5. Quark propagators are calculated with exponentially
smeared sources and local sinks.  We combine quark propagators to form
mesons of the form $SS$, $SV$ and $VV$ and baryons of the form $SSS$,
$SSV$, $SVV$ and $VVV$ where $S$ stands for a valence quark with $K_{val} =
K_{sea}$ and $V$ for a valence quark with $K_{val} \neq K_{sea}$. These
combinations include all unequal quark mass combinations allowed for
degenerate $u$ and $d$ sea quarks and valence $s$ quark.

\begin{figure*}[tb]
\vspace{-5mm}
\begin{center}
\leavevmode
\hspace*{-2mm}
\epsfxsize=7.5cm
\epsfbox{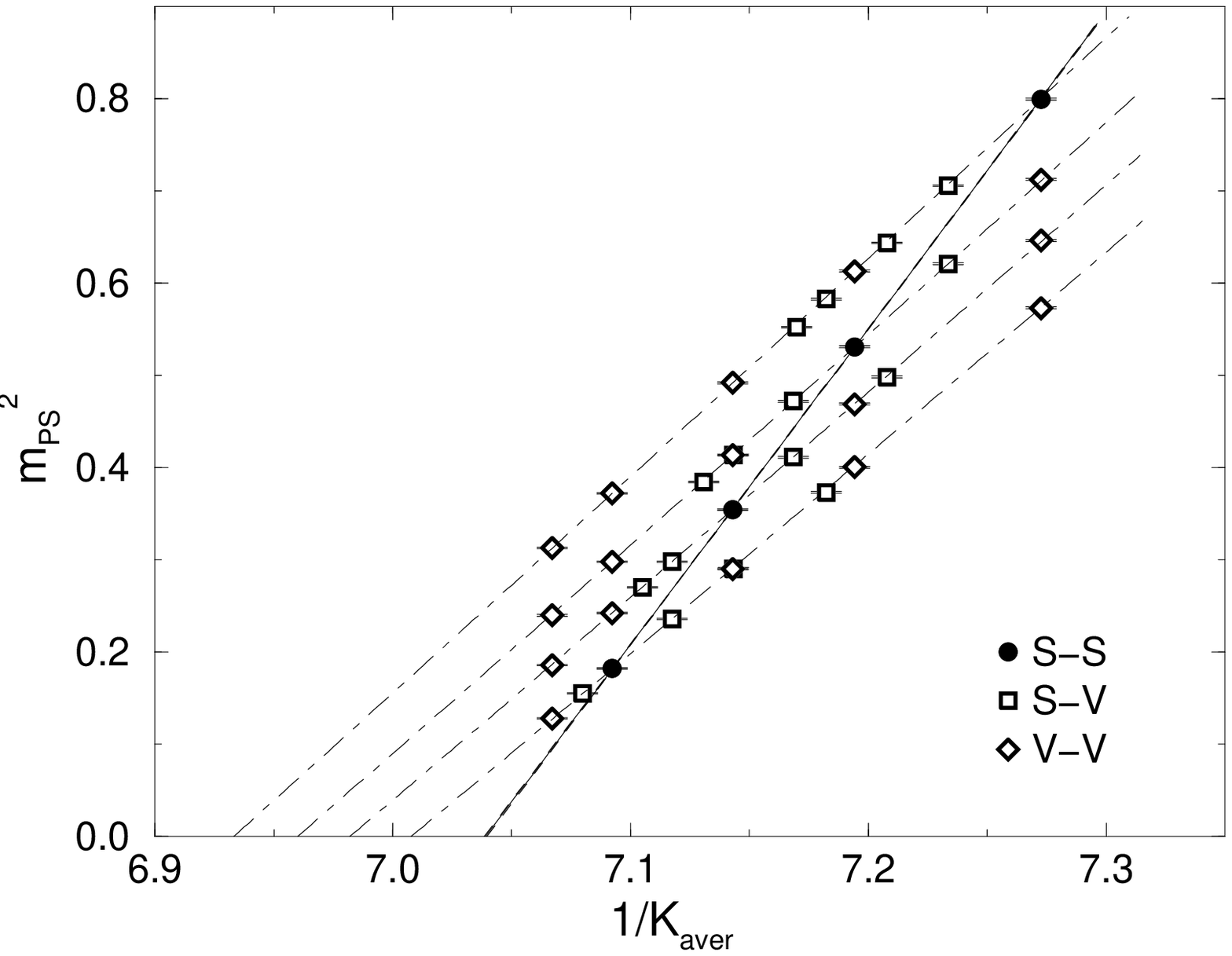}
\hspace*{+8mm}
\epsfxsize=7.5cm
\epsfbox{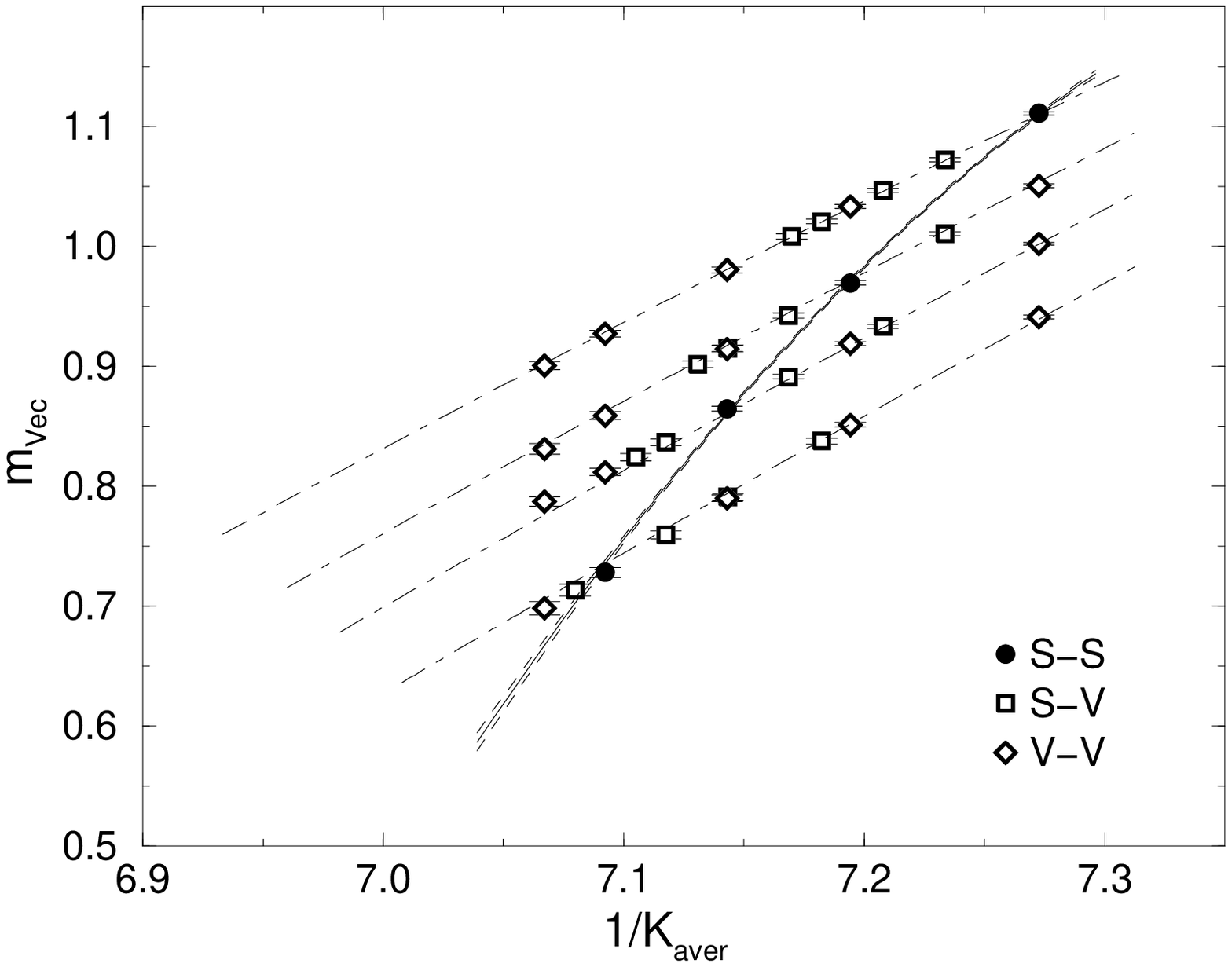}
\end{center}
\vspace{-13mm}
\caption{
Meson masses at $\beta\!=\!1.95$. 
S and V are for valence quarks equal or different
to the sea quark. Lines are results of combined quadratic fits.
}
\label{fig:chiral-full}
\vspace{+2pt}
\end{figure*}

We measure hadron propagators at every 5th trajectory.  Hadron masses are
extracted by uncorrelated fits. Errors are determined with the binned
jackknife method.  The error reaches an approximate plateau after a bin
size of at most 10 configurations (50 HMC trajectories), which we choose
for error estimates of all our runs.

\section{Full QCD Light Hadron Spectrum}

Figure \ref{fig:chiral-full} shows meson masses at $\beta\!=\!1.95$ as a
function of the average hopping parameter
$1/K_{aver}=(1/K_{val(1)}+1/K_{val(2)})/2$ of the two valence quarks in the
meson. Partially quenched data ($VV$ and $SV$) lie along clearly distinct
lines for each $K_{sea}$. Each of these mass results are almost linear for
both pseudoscalar and vector mesons, but their slope shows a variation with
$K_{sea}$. With decreasing sea quark mass we observe a decreasing slope for
pseudoscalar mesons and an increasing slope for vector mesons. 
While the pseudoscalar meson mass in the $SS$ channel is almost linear 
in $1/K_{sea}$, the vector meson mass exhibits a clear concave curvature.

The picture for decuplet baryons looks very similar to the one for vector
mesons, whereas for octet baryons partially quenched data of different
degeneracies lie on clearly different lines.

For chiral extrapolations we fit
at each value of $\beta$ all the data of a given channel
together in a combined fit with a general quadratic ansatz in $1/K_{sea}$
and $1/K_{val}$. 
For pseudoscalar mesons consisting of valence quarks
``1'' and ``2'', the fit ansatz is
\begin{eqnarray}
m_{PS}^2 & \!\!\!\!=\!\!\!\! & B_s m_{sea} + B_v \overline{m}_{val} 
 + C_s m_{sea}^2
 + C_v \overline{m}_{val}^2   \nonumber \\
 &   &   \!\! + C_{sv} m_{sea} \overline{m}_{val}    
 + C_{12} m_{val(1)} m_{val(2)}
\label{eq:ps-fit-full} 
\end{eqnarray}
where bare quark masses are defined as
\begin{eqnarray}
m_{sea}    & = & (1/K_{sea} - 1/K_c)/2 \label{eq:seaquark-full} \\
m_{val(i)} & = &(1/K_{val(i)} - 1/K_c)/2 \label{eq:valquark-full}
\end{eqnarray}
and 
$\overline{m}_{val}$ is the average valence quark mass.
The valence-valence cross term $C_{12}$ is introduced to take into account 
mass differences of mesons at the same average valence quark mass.
For vector mesons and decuplet baryons such a term is dropped because we
observe them to be well described as a function of the sea quark mass and the 
average valence quark mass only.  Curves of the combined fit are illustrated 
in Fig.~\ref{fig:chiral-full} for meson masses. 

For octet baryons we employ a combined quadratic fit for $\Sigma$-like and
$\Lambda$-like baryons simultaneously. Linear terms are taken in a form
inspired by chiral perturbation theory and general quadratic terms of
individual $m_{val(i)}$ are added. The resulting ansatz has 12 parameters
which fits the 84 octet baryon mass data per $\beta$ value well.

\begin{figure}[tb]
\vspace{-2mm}
\centerline{\epsfxsize=7.5cm \epsfbox{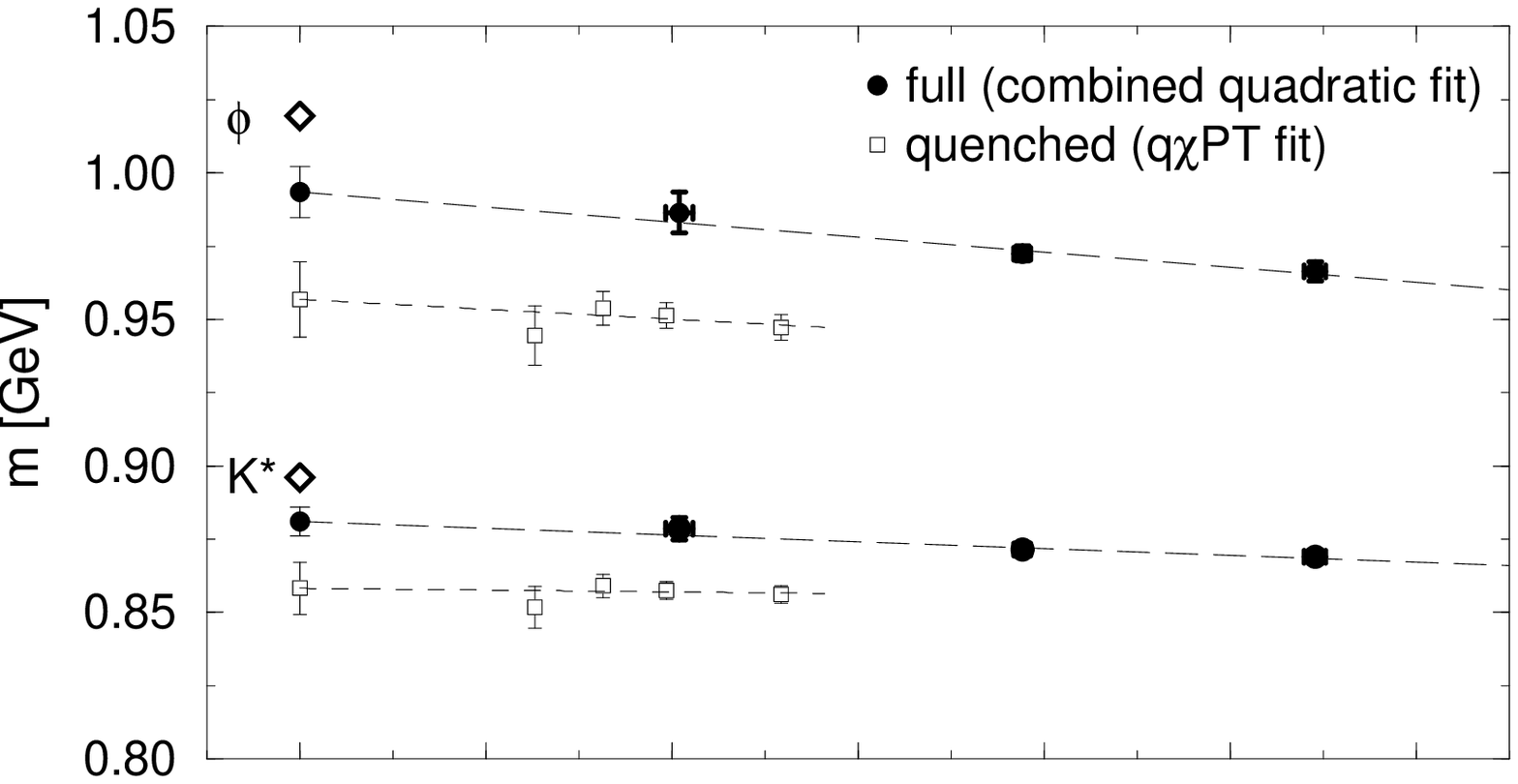}}
\vspace{-7.2mm}
\centerline{\epsfxsize=7.5cm \epsfbox{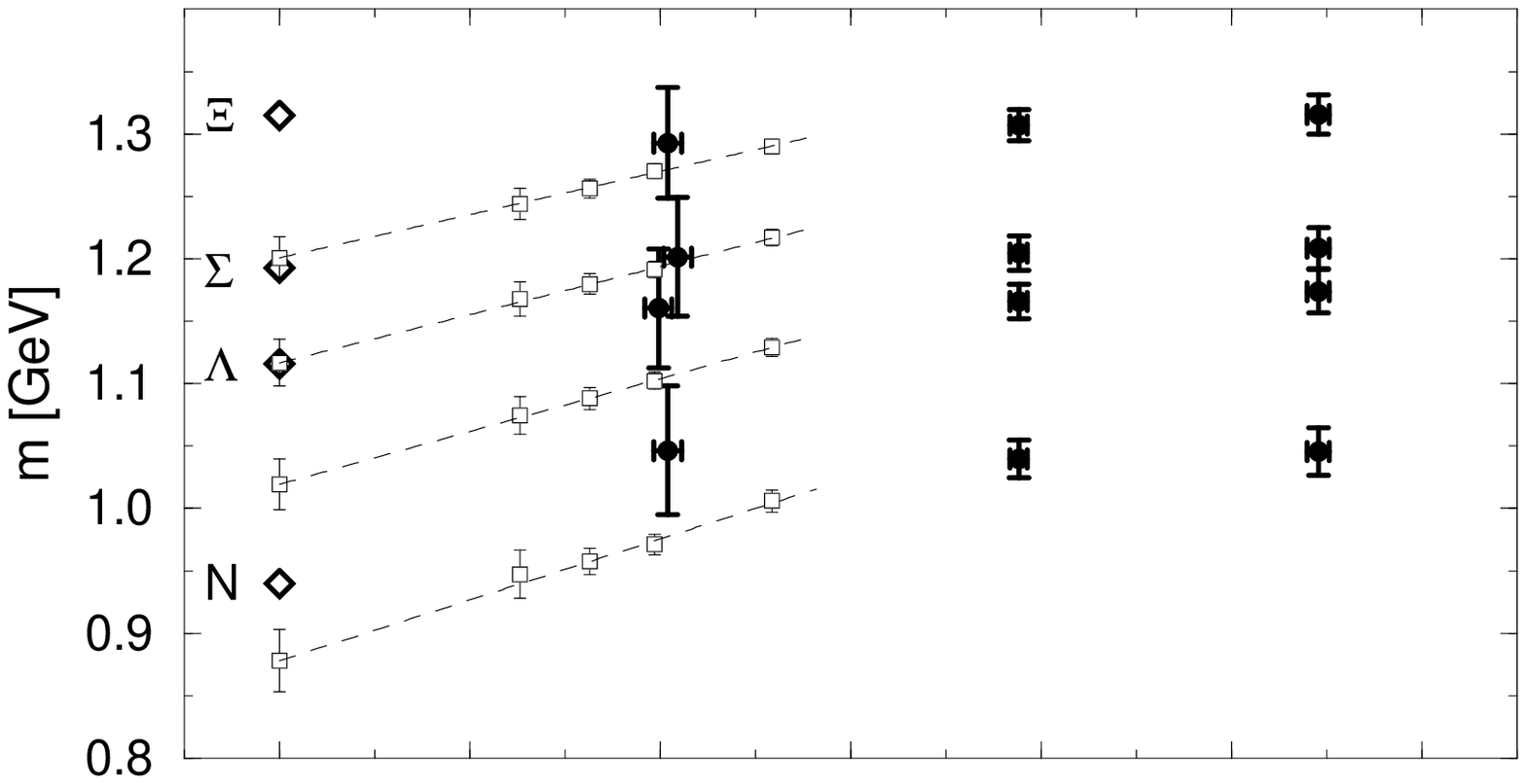}} 
\vspace{-7.2mm}
\centerline{\epsfxsize=7.5cm \epsfbox{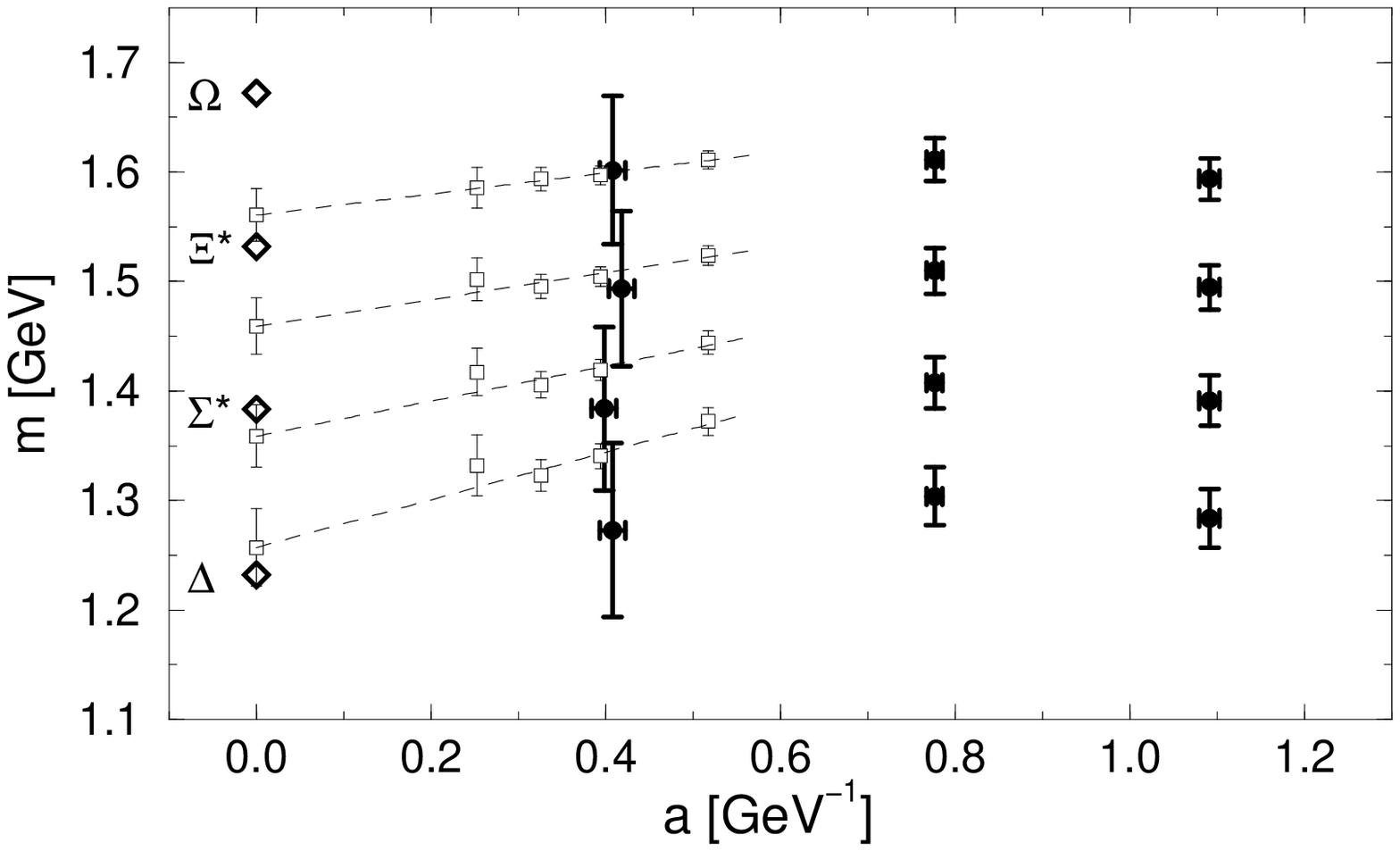}} 
\vspace{-11.2mm}
\caption{Hadron masses in full QCD as function of the lattice spacing.
$M_K$ is used as input.
At the smallest lattice spacing $\Lambda$, $\Sigma$, $\Sigma^*$ and $\Xi^*$
masses are shifted in $x$-direction for better visibility.
For comparison quenched QCD results with the Wilson quark action are
included.} 
\label{fig:cont-full}
\vspace{-15pt}
\end{figure}

Using the results from the combined quadratic fits, the physical light
quark hopping parameter $K_{ud}$ and the lattice scale $a^{-1}$ are 
fixed with the experimental value of $M_\pi$ and $M_\rho$ as input. 
The valence strange quark hopping parameter
$K_{s}$ (at $K_{sea}=K_{ud}$) is determined from either $M_K$
or $M_\phi$.
The light hadron spectrum is calculated by setting $K_{sea}$ to
$K_{ud}$, and $K_{val}$ to either $K_{ud}$ or $K_{s}$ depending on
the quark content of the particle.
The resulting hadron masses as a function of the lattice spacing are shown
in Fig.~\ref{fig:cont-full}. 
In the next section we discuss these results, focussing on
what extent dynamical sea quark effects can be seen in comparison to 
the results from our simulation of quenched
QCD with the standard action, plotted in Fig.~\ref{fig:cont-full}
by open symbols.

\section{Sea Quark Effects}

\subsection{effects on the light hadron spectrum} 

An interesting result which stands out in Fig.~\ref{fig:cont-full} is that 
the $K^*$ and $\phi$ masses from the full QCD simulation lie higher than those
of the quenched results at finite lattice spacing, and that they extrapolate 
to values in the continuum limit considerably closer to experiment. 
If one makes a linear continuum extrapolation as shown by dashed lines, 
the difference from the experimental value is decreased by about 60\% 
compared to quenched QCD.
The expected leading scaling violation for the tadpole-improved SW quark
action is $O(g^2(a)a)$.  We also try such an extrapolation, which yields
similar results.

The origin of the remaining difference from experiment is an interesting
issue.  One possible reason is that sea quark masses explored in the
simulation ($0.8\gsim m_{PS}/m_V\gsim 0.6$) are still relatively heavy so
that sea quark effects are only partly operative.  We should also remember
that our simulation treats only two quark flavours as dynamical.
Therefore, another possibility is that at least part of the difference
reflects the quenched treatment of the strange quark.

It is interesting to see in detail how the difference between quenched and
full QCD arises as a function of the sea quark mass.  
The $J$ parameter 
$J=m_V \, dm_{V} / dm_{PS}^2$
at $m_V/m_{PS}=1.8$ \cite{ref:J-parameter} 
is a measure of the meson hyperfine splitting around the strange quark.
With the strange quark treated in the partially quenched approximation the
derivative is evaluated at fixed sea quark mass by varying the
valence quark mass only. Decreasing the sea quark from a large value down to
the light quark mass interpolates between quenched and two-flavour full QCD.  

\begin{figure}[tb]
\vspace{-10mm}
\begin{center}
\leavevmode
\hspace*{-1mm}
\epsfxsize=7.5cm
\epsfbox{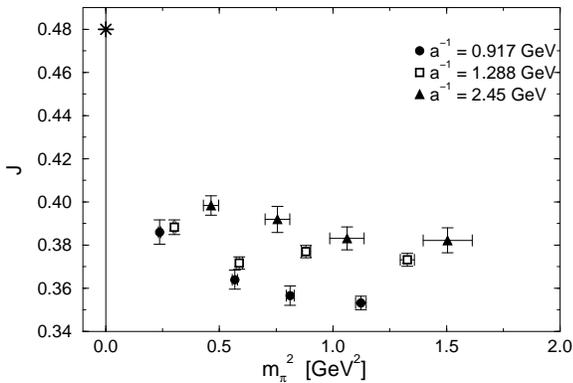}
\end{center}
\vspace{-12mm}
\caption{$J$ parameter at different sea quark masses and lattice
spacings. The experimental value is denoted in the chiral limit with a
star.} 
\label{fig:J-full}
\vspace{-15pt}
\end{figure}

In Fig.~\ref{fig:J-full} we plot $J$ as a function of the sea
quark mass (in the plot the mass squared of the pseudoscalar meson
consisting of two sea quarks is used).
We find the values to be approximately constant for moderate or large sea
quark masses and roughly consistent with the quenched value 
0.346(22) in the limit of infinitely heavy sea quarks. Only when the sea
quark mass becomes close to the chiral limit the value of $J$ rises toward the
experimental value of 0.48, resulting in an increase of the hyperfine
splitting.

Sea quark effects on baryons are less clear. 
While scaling violation appears small for baryon mass results, the errors
are still too large to reliably discuss continuum extrapolations and how
the two-flavour results compare with those of quenched QCD.
Furthermore there are two points that have to be kept in mind: 
(i) The spatial lattice size at the smallest lattice spacing turned out to 
be 1.9~fm compared to 2.5--2.6~fm at the two coarser lattice spacings. 
These are smaller than the size of 3~fm for our quenched simulation, and 
raises the question of finite-size effects.  
(ii) The smallest valence quark mass 
for full QCD corresponds to $m_{PS}/m_{V}\approx 0.5$ rather than $\approx
0.4$ for quenched QCD, possibly leading to 
systematic errors in chiral extrapolations of baryon masses.
Further work matching these points is needed to clarify 
sea quark effects in the baryon mass spectrum.

\subsection{static quark potential}

In full QCD the string between a static quark and antiquark can break
and one would expect to observe a flattening of the potential at large
distances. 

\begin{figure}[tb]
\vspace{-4mm}
\begin{center}
\leavevmode
\hspace*{-1mm}
\epsfxsize=7.3cm
\epsfbox{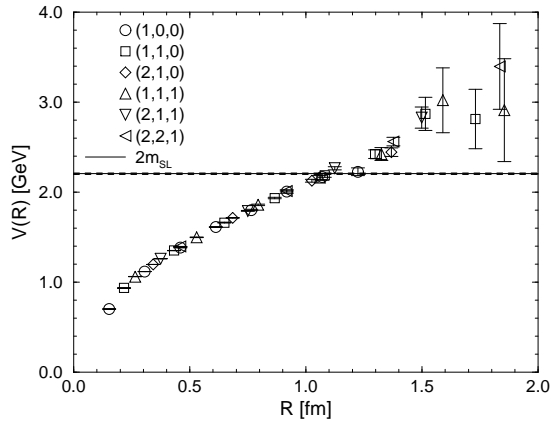}
\end{center}
\vspace{-13mm}
\caption{Static quark potential and static-light meson mass at $\beta =
1.95$ and the lightest sea quark mass.} 
\label{fig:Potential-full}
\vspace{-15pt}
\end{figure}

We extract the value of the potential between static quarks $V(R)$ from
smeared Wilson loops. At $\beta\!=\!1.95$ we also measure the mass of the
static-light pseudoscalar meson $m_{SL}$ with the light quark mass
equal to the sea quark mass. In Fig.~\ref{fig:Potential-full} these two
quantities are compared for the lightest sea quark mass at $m_{PS}/m_{V}
\approx 0.6$. The linear increase of the potential continues beyond the
crossing point with $2m_{SL}$ at around 1 fm. A similar situation is found
also at other sea quark masses. We therefore see no indication for
string breaking, similar to the findings of previous lattice
simulations\cite{ref:guesken}.

A possible explanation of this result is that the Wilson loop does not have
a sufficient overlap with the state of a broken string.
In \cite{ref:kaneko} we argue, that in order to find the
true ground state one would have to extract the potential from Wilson loops
at a large time separation. This is difficult in practice because of the
rapidly vanishing signal in the noise. One would have to use (an)
operator(s) with a better overlap with a static-light meson pair
state. This is also the essence of other recent proposals\cite{ref:string}.

A quantity in which clear sea quark effects can be seen is the Coulomb
coefficient $\alpha$. Here the full QCD results are larger than quenched,
and they increase with decreasing sea quark mass. The magnitude of the
shift from quenched to full QCD is consistent with expectations for
$N_f\!=\!2$~\cite{ref:kaneko}.

\section{Light Quark Masses in Full QCD}

It has been known for some time that light quark masses calculated in 
two-flavour full QCD take significantly smaller values compared to 
those in quenched QCD\cite{ref:ukawa}. 
Most recently the SESAM collaboration\cite{ref:SESAMQuark} reported
$m_{ud}\!=\!2.7(2)$~MeV 
from their simulation with the standard action at $a^{-1}\!\approx\!2.5$ GeV 
($\beta\!=\!5.6$). They furthermore pointed out that the ratio $m_s/m_{ud}$ 
takes a large value $\sim 50$ in contrast to the value $\sim 25$ commonly 
quoted from chiral perturbation theory. 

These results have been obtained with the VWI definition of the quark mass
$m_q^{\rm VWI} \!=\! Z_m (1/K_q \!-\! 1/K_{crit})/2$, where
$K_c^{sea}$, the hopping parameter where the PS meson mass vanishes when 
valence and sea quarks have the same mass, is substituted for $K_{crit}$. 

It has been suggested\cite{ref:Gupta} that the use of the partially 
quenched critical value $K_c^{\rm PQ}$, 
defined by $m_{PS}(K_{val})\!=\!0$ with $K_{sea} \!=\! K_{ud}$ fixed,
resolves the problems described above. 
In two-flavour QCD however, the choice $K_c^{sea}$ is the most natural one for 
the average $u$ and $d$ quark mass $m_{ud}$, since the sea quarks 
are identified with the degenerate $u$ and $d$ quark.  
Furthermore, $K_c^{sea}$ and $K_c^{\rm PQ}$ coincide in the continuum limit. 
If the difference between them vanishes sufficiently fast when approaching
the continuum limit, then also the two $m_q^{\rm VWI}$ will agree.  
Indeed, calculating $m_q^{\rm VWI}$ in perturbation theory, we find that
the difference between the two quark mass definitions vanishes as $O
( g^4a\log a )$. 

\begin{figure}[tb]
\vspace{-2mm}
\centerline{\epsfxsize=8cm \epsfbox{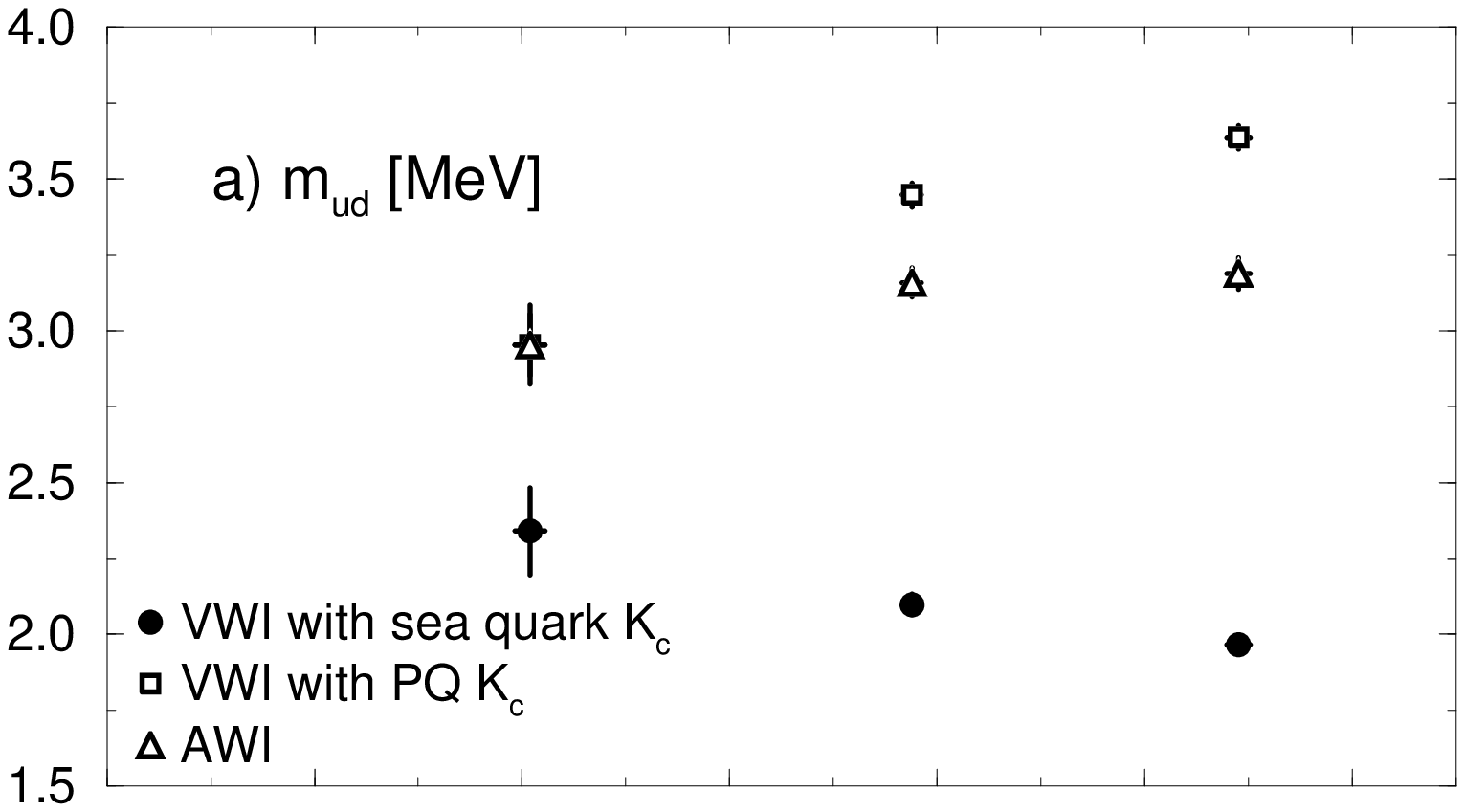}}
\vspace{-6mm}
\centerline{\epsfxsize=8cm \epsfbox{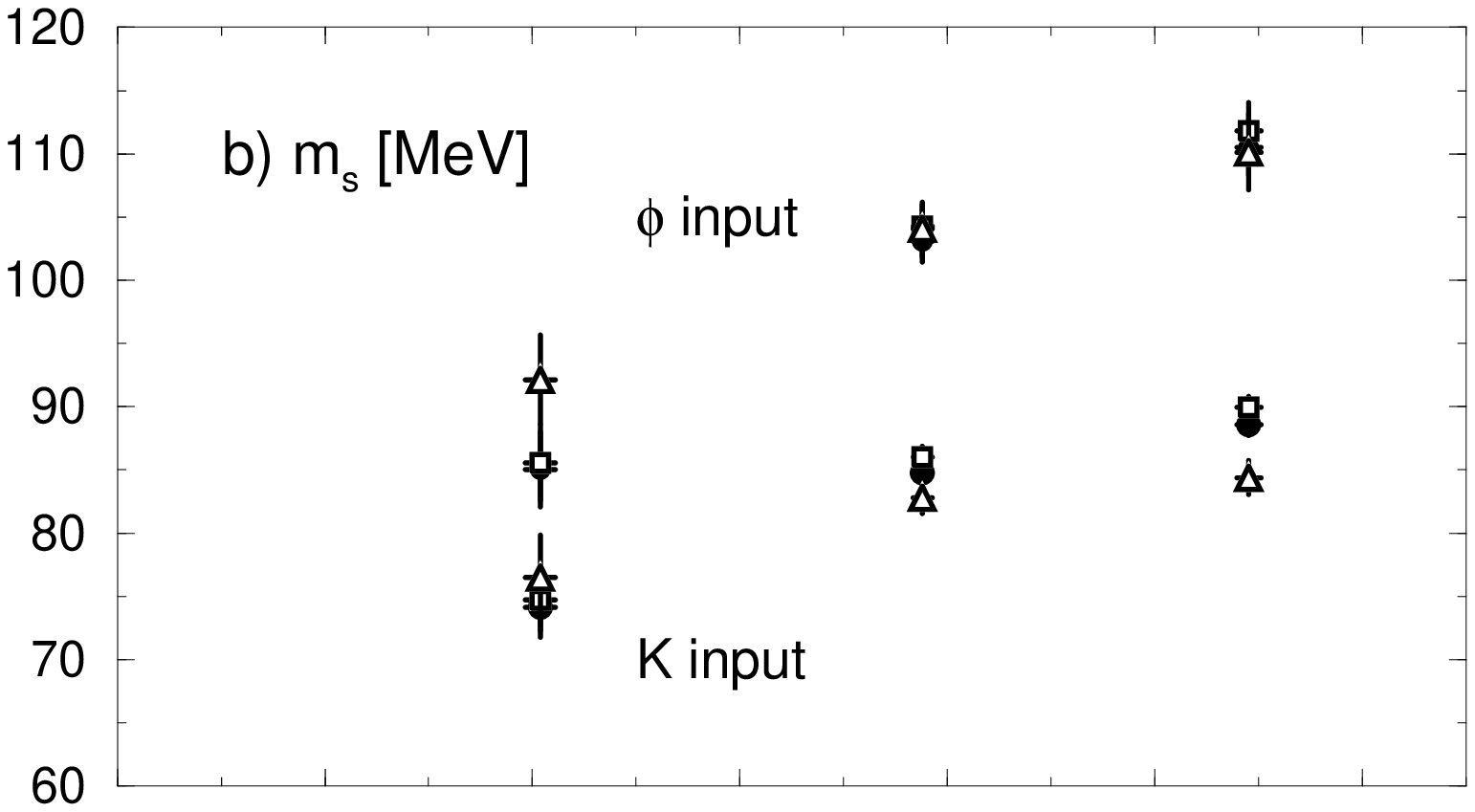}} 
\vspace{-6mm}
\centerline{\epsfxsize=8cm \epsfbox{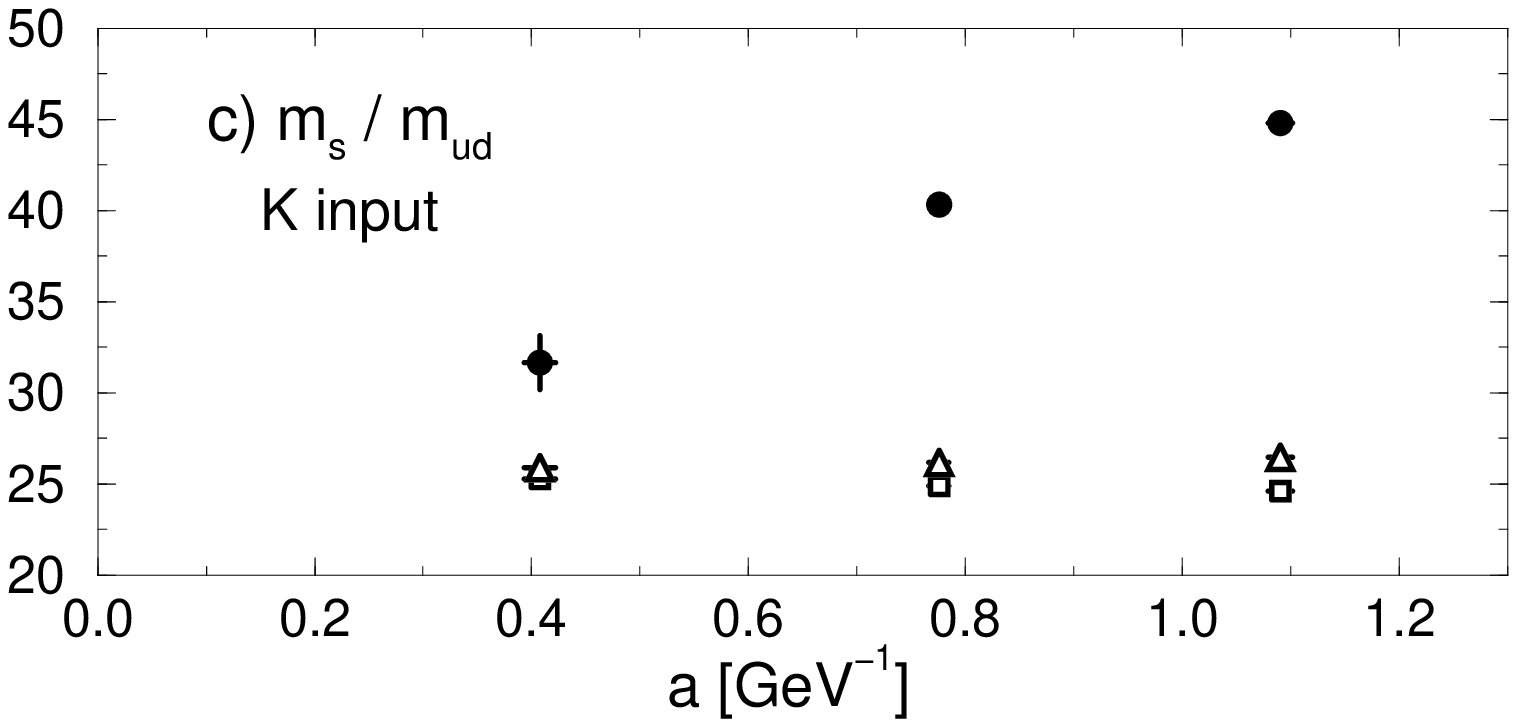}} 
\vspace{-10mm}
\caption{Quark mass in full QCD with various definitions.}
\label{fig:quark-full}
\vspace{-15pt}
\end{figure}

Figure \ref{fig:quark-full}a shows our results for $m_{ud}$ in the 
$\overline{\rm MS}$ scheme at $\mu\!=\!2$~GeV, calculated with
tadpole-improved one-loop  
values of renormalization factors\cite{ref:taniguchi}, 
as a function of the lattice spacing. 
The two definitions of $m_q^{\rm VWI}$ lead to significantly different
values at finite lattice spacings but as expected they converge towards the
continuum limit.  In addition the AWI quark mass also extrapolates
to a similar value.
The value reached is 2.5--3 MeV, which is about 40\% smaller than the result
obtained in quenched QCD. A theoretical argument for a decrease of quark 
masses by unquenching has been previously made from different running of 
the $\beta$ function\cite{ref:Mackenzie}.

Results for the strange quark mass are shown in Fig.~\ref{fig:quark-full}b. 
The difference between various quark mass definitions is much smaller than
for the light quark.  The main uncertainty stems from the choice of the
experimental input to fix $K_{s}$.  We find values in the
continuum limit of about $m_s \approx 70$ MeV ($M_K$ input) or 80 MeV
($M_{\phi}$ input). The discrepancy between these two values is
smaller than in quenched QCD. The remaining difference might be
due to the quenching of the strange quark or 
that the sea quarks are too heavy.

In Fig.~\ref{fig:quark-full}c we show the ratio of strange to light quark
masses as function of the
lattice spacing. While for VWI masses with the choice
$K_{crit}\!=\!K_c^{sea}$ the value is $\sim\! 45$ at the largest lattice
spacing, it decreases toward the continuum limit and all definitions
converge to $m_{s}/m_{ud} \approx 25$, consistent with \cpt results.

\section{Conclusions}

In our extensive study of the quenched hadron mass spectrum, we found that 
the mass results show an overall consistency with \qcpt predictions, 
which led us to employ \qcpt mass formulae for chiral extrapolations 
in our final analysis. The spectrum in the continuum limit
deviate systematically and unambiguously from experiment, on a
level of about 10\% (5$\sigma$).  The terms specific to q$\chi$PT, however, 
have a small magnitude of 2--3\% of hadron masses, and are hence responsible 
only for a small part of the deviation.   
While chiral extrapolations differ, our main conclusion remains unchanged 
since Lattice 97.

Our study of full QCD is in a less advanced stage, attempting to
systematically approach a realistic QCD simulation. Still, already several
intriguing results are found. The discrepancy in the meson hyperfine
splitting is reduced by about 60\%  and light quark masses are found to be
around 40\% smaller than in quenched QCD.
In the future it will be interesting to look for sea quark effects also in
other areas such as topology and heavy quark physics.  Work in these
directions has already started.

\vspace{8pt}

I would like to thank the members of the CP-PACS collaboration for their
continuous help and support during the preparation of this overview. 
This work is in part supported by the Grant-in-Aid No. 09304029 of the
Ministry of Education.

\end{document}